\documentclass[prl,aps,twocolumn,floats,showpacs]{revtex4}
\usepackage{bm,graphicx}

\newcommand{\be}{\begin{equation}}
\newcommand{\ee}{\end{equation}}
\newcommand{\bea}{\begin{eqnarray}}
\newcommand{\eea}{\end{eqnarray}}
\newcommand{\phrl}[1]{Phys.~Rev.~Lett. {\bf #1}}
\newcommand{\phrb}[1]{Phys.~Rev.~B {\bf #1}}

\newcommand{\bib}{\bibitem}
\newcommand{\lb}{\left[}
\newcommand{\rb}{\right]}
\newcommand{\lp}{\left(}
\newcommand{\rp}{\right)}

\renewcommand{\k}{{\bf k}}
\newcommand{\p}{{\bf p}}

\newcommand{\ua}{\uparrow}
\newcommand{\da}{\downarrow}
\newcommand{\F}{{\cal F}}

\begin{document}

\title{Spin analog of the controlled Josephson charge current}
\author{Sudhansu S. Mandal and Soumya P. Mukherjee}
\affiliation{Theoretical Physics Department, Indian Association for the 
   Cultivation of Science, Jadavpur, Kolkata 700 032, India}

\date{\today}


\begin{abstract}
We propose a controlled Josephson spin current across the
junction of two non-centrosymmetric superconductors like CePt$_3$Si.
The Josephson spin current arises due to direction
dependent tunneling matrix element and different momentum dependent
phases of the triplet components of the gap function.
Its modulation with the
 angle $\xi$ between the noncentrosymmetric axes of 
two superconductors is proportional to $\sin \xi$.  
This particular dependence on $\xi$ may find application of the proposed set-up
in making a {\it Josephson spin switch}.

\end{abstract}

\maketitle


Traditionally Josephson junctions \cite{Barone,Golubov} in superconductors 
draw interest both scientifically and its applicability in making devices.
With no exception, they have also been studied in unconventional superconductors 
like spin-singlet cuprate \cite{Tsuei}
and spin-triplet Sr$_2$RuO$_4$ \cite{Mackenzie} superconductors. However no
Josephson junction between nonmagnetic superconductors is known
to generate spin-polarized current.
The purpose of this letter is to theoretically show that the 
direction dependent tunneling matrix element
across the junction between two recently discovered
\cite{Bauer} non-centrosymmetric superconductors like
CePt$_3$Si, leads to tunneling of both spin-singlet and spin-triplet Cooper
pairs. As a consequence, nonvanishing spin-Josephson current is viable along
with the usual charge-Josephson current. This {\it novel} spin-Josephson
current depends on the relative angle $\xi$ between the axes of 
non-centrosymmetry $\hat{n}_L$ and $\hat{n}_R$ in the left and right
side of the junction respectively. This angular dependence may be used to
make a {\it Josephson spin switch.}

The normal state Hamiltonian \cite{Gorkov,Sigrist}
for the electrons in a band of 
a lattice without inversion symmetry is
\be
  H_0 = \sum_{\k, s}\xi_\k c_{\k s}^\dagger c_{\k s} + \sum_{\k,s,s'}
       \bm{g}_\k \cdot \bm{\sigma}_{ss'}c_{\k s}^\dagger c_{\k s'} \, ,
\label{H_normal}
\ee
where electrons with momentum $\k$ and spin $s\, (=\uparrow \text{or}
\downarrow)$ are created (annihilated) by
the operators $c_{\k s}^\dagger$ ($c_{\k s}$), and
$\xi_\k$ is the band energy measured from the Fermi energy $\epsilon_F$. 
The second term in the Hamiltonian
(\ref{H_normal}) breaks parity symmetry as $\bm{g}_{-\k} = -\bm{g}_\k$ for a
non-centrosymmetric system. For a system like Heavy fermion compound CePt$_3$Si
which has layered structure, $H_0$ is considered to be two-dimensional. For such a
system of electrons with band mass $m$, $\xi_\k = \frac{\k^2}{2m} -\epsilon_F$
and $\bm{g}_\k = \alpha \bm{\eta}_\k$ where $\bm{\eta}_k = \hat{n} \times \k$, 
i.e., the spin-orbit interaction
is of Rashba type \cite{Rashba} and $\alpha$ is the Rashba parameter. Here  
$\hat{n}$ represents the axis of non-centrosymmetry which is perpendicular to
the plane of the system.  
Due to the breaking down of the parity, spin
degeneracy of the band is lifted; by diagonalizing $H_0$,
one finds two spin-split bands with energies $\xi_{\k\lambda} = \xi_k
+\lambda \alpha \vert \k\vert$ where $\lambda = \pm$ describes
helicity of the spin-split bands. Therefore in the 
diagonalized basis $H_0$ (\ref{H_normal}) becomes
$H_0 = \sum_{\k , \lambda = \pm} \xi_{\k\lambda }\tilde{c}_{\k \lambda}^\dagger
\tilde{c}_{\k \lambda} $,
where $\tilde{c}_{\k \lambda} =\lp c_{\k \uparrow} - i \lambda \exp \, 
(i \phi_\k ) c_{\k \downarrow} \rp /\sqrt{2}$ is the electron destruction operator
and $\tilde{c}_{\k \lambda}^\dagger =\lp c_{\k \uparrow}^\dagger
 + i \lambda \exp \, (-i
\phi_\k ) c_{\k \downarrow}^\dagger \rp /\sqrt{2}$ is the electron creation
operator in band $\lambda$ with momentum $\k$ whose orientation 
with $\hat{x}$-axis is $\phi_\k$. 
The density of electronic states at Fermi energy in these bands may be found as
$\nu_\lambda = \frac{m}{2\pi}\lp 1 - \lambda 
m\alpha / \sqrt{k_F^2+m^2\alpha^2} \rp$, where $k_F= \sqrt{2m\epsilon_F}$ is
the Fermi momentum.

Band structure calculation \cite{Bose} on CePt$_3$Si reveals that the energy difference
between two spin-split bands near $k_F$ is 50--200 meV which is much larger 
than the superconducting critical temperature, $k_BT_c \approx 0.06$ meV 
\cite{Bauer}. 
The formation of Cooper pairing between electrons in different
spin-split bands may thus be ignored. 
The Hamiltonian for these superconductors
may then be written as
\be
H_1 = \sum_{\k , \lambda = \pm} \lb \xi_{\k\lambda }\tilde{c}_{\k \lambda}^\dagger
\tilde{c}_{\k \lambda} +\lp \Delta_{\k\lambda} 
\tilde{c}_{\k \lambda}^\dagger \tilde{c}_{-\k \lambda}^\dagger + \, \rm{h. c.}
 \rp \rb  \, ,
\label{H_super}
\ee
where Cooper pairs are only between intraband electrons. 
Therefore the normal and anomalous Green's functions are obtained 
respectively as
${\cal G}_\lambda (\k , i \epsilon_n)=-(i \epsilon_n +\xi_{\k\lambda})/
(\epsilon_n^2+\xi_{\k\lambda}^2+\vert \Delta_{\k\lambda}\vert^2)$ and
${\cal F}_\lambda (\k , i \epsilon_n)=\Delta_{\k\lambda} /
(\epsilon_n^2+\xi_{\k\lambda}^2+\vert \Delta_{\k\lambda}\vert^2)$, where
$\epsilon_n$ is the fermionic Matsubara frequency \cite{Mahan}.
The superconducting
order parameter $\Delta_{\k\lambda}$ obeys the symmetry \cite{Bose,Curnoe}: 
$\Delta_{-\k\lambda} = -\Delta_{\k\lambda}$. 
We consider $\Delta_{\k\lambda}
= \tilde{\Delta}_{k\lambda} \Lambda_\k$, i.e., the angular dependence 
of $\k$ on the order parameters of two bands are assumed to be same.
Apart from the overall phase rigidity angle $\Theta$ of the
superconductor, there may be a relative phase difference $\theta$
between superconductors of two bands: $\tilde{\Delta}_{k+} = \vert
 \tilde{\Delta}_{k+} \vert e^{i \Theta}$ and
 $\tilde{\Delta}_{k-} = \vert \tilde{\Delta}_{k-} \vert e^{i (\Theta + \theta)}$. 
Reverting $H_1$ (\ref{H_super}) to spin-up (-down) basis, we find
\be
H_1 = \sum_\k \Psi_\k^\dagger \lp \begin{array}{cccc}
  \xi_\k & \Delta_{\k, \uparrow\uparrow} &
  \Gamma_{\k ,\rm{R}} & \Delta_{\k , \uparrow\downarrow}\\

 \Delta^*_{\k, \uparrow\uparrow} & -\xi_\k &
 \Delta^*_{\k, \downarrow\uparrow}& \Gamma^*_{\k ,\rm{R}}  \\

 \Gamma^*_{\k ,\rm{R}} & \Delta_{\k ,\downarrow\uparrow} &
  \xi_\k & \Delta_{\k , \downarrow\downarrow} \\

\Delta_{\k , \uparrow\downarrow}^* & \Gamma_{\k , \rm{R}} & 
\Delta_{\k , \downarrow\downarrow}^* & -\xi_\k 
\end{array} \rp \Psi_\k  \, ,
\label{H_trans}
\ee
where $\Delta_{\k , \uparrow\uparrow}=
\frac{1}{2}(\tilde{\Delta}_{k+} +\tilde{\Delta}_{k-})\Lambda_\k$, $\Delta_{\k ,\downarrow
\downarrow}= \frac{1}{2}\exp \, [2i 
\phi_\k](\tilde{\Delta}_{k+}+\tilde{\Delta}_{k-})\Lambda_\k$, and $\Delta_{\k , \uparrow
\downarrow} = -\Delta_{\k , \downarrow\uparrow} = 
\frac{i}{2}\exp\, [i \phi_\k](\tilde{\Delta}_{k+}-
\tilde{\Delta}_{k-})\Lambda_\k$  
are the different components of pairing potential $\Delta_{\k , ss'}$ between
electrons with spins $s$ and $s'$, and $\Gamma_{\k , \rm{R}} = i
\alpha \vert \k \vert \exp \, [-i\phi_\k]$ is the Rashba spin-orbit
coupling potential.  The Hamiltonian 
(\ref{H_trans}) has been expressed in the basis such that
 $\Psi_\k^\dagger = (c_{\k\uparrow}^\dagger ,\, c_{-\k\uparrow},\,
 c_{\k \downarrow}^\dagger , \, c_{-\k\downarrow})$.
Since $\Delta_{\k , \ua\da} = -\Delta_{\k , \da\ua}$, there is no triplet
component with zero projection along the spin-quantization direction; 
pairing between electrons with unequal spins
entirely gives rise to singlet component. 
If we choose $\Lambda_\k = -i\exp \, [-i \phi_\k]$, the triplet component of
the pairing may be expressed as $\hat{\Delta}_{\k ,\rm{T}}=
(\bm{d}_\k\cdot\bm{\sigma})
i \sigma_y$ with $\bm{d}_\k = 
\frac{1}{2\vert \k\vert}(\tilde{\Delta}_{k+} +\tilde{\Delta}_{k-})
\bm{\eta}_\k$. 
In other words, the only stable \cite{Sigrist} spin triplet component   
whose $\bm{d}_\k \parallel \bm{\eta}_\k$
corresponds to this $\Lambda_\k$. 
Therefore $\Delta_{\k ,\ua\ua}$ and $\Delta_{\k , \da\da}$
will have equal but opposite momentum-dependent phase.
 Accordingly the singlet component of the pairing potential
becomes $\Delta_{\k , \rm{S}} = \frac{1}{2}(\tilde{\Delta}_{k+}
 -\tilde{\Delta}_{k-})$.
If $\vert \tilde{\Delta}_{k+}\vert \neq \vert \tilde{\Delta}_{k-}\vert$,
the admixture \cite{Gorkov,Sigrist}
of singlet and triplet pairing takes pace.
Recent observation \cite{Expt} 
of Josephson current in the junction of CePt$_3$Si and
$s$-wave superconductor suggests the existence of spin-singlet
order parameter, while larger \cite{Bauer} upper critical field $ H_{c2}$
seems to suggest that the spin-triplet pairing occurs. 
One normally assumes the superconducting order parameter to be independent
of the magnitude of momentum, i.e., $\tilde{\Delta}_{k\lambda}$ is
$k$-independent. 
The anomalous Green's function ${\cal F}_{ss'}(\k ,i\epsilon_n)$
in the up-down basis is related to ${\cal F}_\lambda$ as
$\F_{\ua\ua} = \frac{1}{2}(\F_{+}+\F_{-})$,
$\F_{\da\da} = \frac{1}{2}e^{2i \phi_\k}(\F_{+}+\F_{-})$, and
$\F_{\ua\da} = -\F_{\da\ua}= \frac{i}{2}e^{i \phi_\k}(\F_{+}-\F_{-})$.
Apart from the applicable momentum-dependent phases, triplet (singlet) 
component of ${\cal F}_{ss'}$ is the addition (subtraction) of anomalous
Green's function of two spin-split bands.

\begin{figure}
\includegraphics[height=6cm,angle=0]{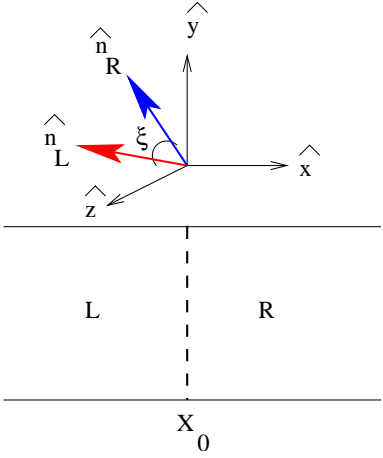}
\caption{Schematic picture of two non-centrosymmetric superconductors,
  denoted by $L$ and $R$, coupled 
   at $x=x_0$. Their orientations are characterized by the axes of 
non-centrosymmetry, $\hat{n}_L$ and $\hat{n}_R$.}
\end{figure}

We now consider Josephson tunneling between two such non-centrosymmetric
superconductors as depicted in Fig.~1. 
The Hamiltonian for the system then reads $H = H_L +H_R
+H_T$, where $H_L$ and $H_R$ are the bulk Hamiltonian of left and right
side of the junction respectively and $H_T$ describes the tunneling between
these two sides. The Hamiltonian $H_L$ is described by $H_1$ (\ref{H_super})
 and $H_R$ is also defined equivalently with the change in the notation of
momentum $\k \to \p$ 
to distinguish each side of the junction. 
Further, each parameter in the left (right) is denoted by 
superscript or subscript L (R).
The tunneling Hamiltonian reads
$H_T = \sum_{\k \p , s} \lp T_{\k\p}c_{\k s}^\dagger c_{\p s} 
  +T^*_{\k\p}c_{\p s}^\dagger c_{\k s} \rp$ , 
where $T_{\k\p}$ is the tunneling matrix element for an electron with momentum
$\p$ to tunnel from right side to left side with momentum $\k$. 
Time-reversal symmetry of $H_T$ suggests $T_{-\k,-\p} = T^*_{\k\p}$.

In a charge-tunneling process, number of electrons in either side of the 
junction changes with time, e.g., the rate of change in the number of electrons
of the right side of the junction $N_R = \sum_{\p ,s} c_{\p s}^\dagger
c_{\p s}$ reads as $\dot{N}_R = i[H_T,\, N_R]$. Therefore the tunneling
charge-current becomes $I_c(t) = -e \left\langle \dot{N}_R (t) \right\rangle$: 
one part corresponds to quasi-particle tunneling and the other part is due
to process of Josephson tunneling \cite{Ambegaokar,Mahan}, i.e., 
tunneling of the Cooper pairs. 
We are herewith interested in the Josephson charge current defined as $I_J^c (t)
= 2e \, \Im\, \lb \exp\, \lp -2ieVt \rp \Phi^c_{ret} (eV)\rb$ 
at time $t$ and for an applied
bias voltage $V$ across the junction, where $\Phi^c (i\omega_m) =-\int_0^\beta
d\tau\, e^{i\omega_m\tau} \left\langle T_\tau A(\tau) A(0) \right\rangle$
with imaginary time $\tau$,
 $A(\tau) = \sum_{\k\p s}T_{\k\p} c_{\k s}^\dagger (\tau) c_{\p s}(\tau')$,
the bosonic Matsubara frequency \cite{Mahan} $\omega_m$,
and inverse temperature $\beta$.
Instead of the axes of noncentrosymmetry of two sides of the junction
being parallel, general axes 
$\hat{n}_L$ and $\hat{n}_R$ will lead to $\hat{\bm{\eta}}_\k = \hat{n}_L \times
\hat{\k}$ and $\hat{\bm{\eta}}_\p = \hat{n}_R \times \hat{\p}$. In that case,
Josephson tunneling occurs in both singlet and triplet channels. Separating
these channels, we find
\be
\Phi^c (i\omega_m)
= - \sum_{i\epsilon_n,\k\p}\frac{\vert T_{\k\p}\vert^2}{2\beta}\lb
\chi_S +(\hat{\bm{\eta}}_k \cdot \hat{\bm{\eta}}_\p) \chi_T \rb \, ,
\ee
where  
\be
\chi_S = \sum_{\lambda ,\lambda'}\lambda\lambda' \Lambda_\k\Lambda_\p^*
\F_\lambda^\dagger (\k , i\epsilon_n)\F_{\lambda'} (\p , i\epsilon_n
  -i\omega_m) 
\ee
and
\be
\chi_T = \sum_{\lambda ,\lambda'} \Lambda_\k\Lambda_\p^*
\F_\lambda^\dagger (\k , i\epsilon_n)\F_{\lambda'} (\p , i\epsilon_n
   -i\omega_m)
\ee
represent singlet and triplet contributions respectively.
While the direction independent electronic tunneling probability $\vert
T_{\k\p}\vert^2 = \vert T\vert^2$ leads to tunneling of singlet 
component only, the triplet components of Cooper pairs tunnel 
due to realistic direction dependent \cite{Duke} $\vert T_{\k\p}\vert^2$.
This triplet part in the charge Josephson tunneling process may effectively
provide spin Josephson tunneling.
The rate of change of spin $S_z = 
\sum_{\p }[c_{\p \ua}^\dagger c_{\p \ua}-c_{\p \da}^\dagger c_{\p \da}]$ 
due to tunneling, i.e., $\dot{S}_z
= i[H_T,\, S_z]$, 
gives rise to spin-Josephson tunneling
current $I_J^s (t) = -2\, \Im [ \exp (-2ieVt) \Phi^s_{{\rm ret}}(eV)]$ where
$\Phi^s (i\omega_m) = -\int_0^\beta d\tau \,e^{i\omega_m\tau} \left\langle T_\tau
B(\tau) A(0) \right\rangle$ with $B(\tau) = \sum_{\k\p } T_{\k\p}\,  
\, [c_{\k \ua}^\dagger (\tau)c_{\p \ua}(\tau)-
c_{\k \da}^\dagger (\tau)c_{\p \da}(\tau)]$.
We thus find
\be
\Phi^s(i\omega_m) = i\sum_{i\epsilon_n ,\k\p}\frac{\vert T_{\k\p} \vert^2}{
2\beta}\sum_j (\hat{\bm{\eta}}_k \times \hat{\bm{\eta}}_\p)_j \chi_T  \, ,
\label{Phis}
\ee
where $ (\hat{\bm{\eta}}_k \times \hat{\bm{\eta}}_\p)_j$ represents $j$-th spatial
component of $\hat{\bm{\eta}}_k \times \hat{\bm{\eta}}_\p$.

We assume that the non-centrosymmetric axes are parallel to the 
interface of the junction
as shown in Fig.~1. The probability of tunneling \cite{Duke}
will be the most along
the direction transverse to the interface. In terms of normal state electronic
tunneling conductance $G_N = I_N/V$ with $I_N$ being normal state tunneling
current, we find dc $(V=0)$ charge and spin Josephson currents:
\be
  I_J^c =  \lp \frac{G_N}{ e {\cal F}}\rp 
 \lb \sin \psi\,  (g_+{\cal A}_1 +g_- {\cal A}_2)
 +  \cos \psi \, (g_+ {\cal A}_3 +g_- {\cal A}_4) \rb 
 \label{I_charge}
\ee
\be
 I_J^s =   \lp \frac{G_N}{e^2 {\cal F}}\rp 
\delta \kappa \sin\xi 
 \lb \cos \psi\, ({\cal A}_1 +{\cal A}_2) 
- \sin \psi \,   ( {\cal A}_3 + {\cal A}_4) \rb \, ,
\label{I_spin}
\ee
where $\psi = \Theta_L - \Theta_R $, 
$g_\pm = \delta\cos \xi \pm 1$, $ \delta \in (0,1)$
is a parameter depending on the model \cite{Matrix}
of tunneling matrix element, i.e., the
angular dependence of $T_{\k\p}$, $\cos \xi = \hat{n}_L \cdot \hat{n}_R$
where both $\hat{n}_L$ and $\hat{n}_R$ are in the plane of the interface,
and $\kappa$ is the projection of $\hat{n}_L \times \hat{n}_R$ along 
perpendicular to the interface. Further
${\cal F} = \pi\sum_{\lambda\lambda'}\nu_\lambda^L \nu_{\lambda'}^R (1+\delta 
\lambda\lambda'\cos \xi )$ describes the dependence of normal state tunneling
current on $\delta$, 
${\cal A}_1 = \Gamma_{++}+\Gamma_{--} \cos (\theta_L -\theta_R)$, 
${\cal A}_2 = \Gamma_{-+}\cos \theta_L +\Gamma_{+-}\cos \theta_R$, 
${\cal A}_3 =  \Gamma_{--} \sin (\theta_L -\theta_R)$, and
${\cal A}_4 = \Gamma_{-+}\sin \theta_L -\Gamma_{+-}\sin \theta_R$ with 
\be
\Gamma_{\lambda\lambda'} = 
\pi \nu_\lambda^L\nu_{\lambda'}^R 
\frac{\vert\tilde{\Delta}_\lambda^L\vert \vert\tilde{\Delta}_{\lambda'}^R\vert} 
{\vert\tilde{\Delta}_\lambda^L\vert + \vert\tilde{\Delta}_{\lambda'}^R\vert} 
 K\lp 
\frac{\vert \tilde{\Delta}_\lambda^L\vert -
 \vert\tilde{\Delta}_{\lambda'}^R\vert } 
{\vert\tilde{\Delta}_\lambda^L\vert + \vert\tilde{\Delta}_{\lambda'}^R\vert} 
 \rp 
\ee
at zero temperature, 
where $K$ is the elliptic function of first kind.

Charge (\ref{I_charge}) and spin (\ref{I_spin}) dc Josephson currents
depend on both sine and cosine of the global phase difference $\psi$
between two superconductors. Instead of continuous
relative phase difference between superconductors in each spin-split band,
there is a possibility of phase locking such that $\theta_L ,\, \theta_R = n\pi$ 
$(n \in Z)$. In this case, ${\cal A}_3 = {\cal A}_4 = 0$ and hence $I_J^c
(V=0) \propto \sin \psi$ 
and $I_J^s (V=0)
\propto \cos \psi $. This situation is also true for $\theta_L
= \theta_R$, even in the absence of their locking at $n\pi$. The critical
charge and spin Josephson currents are modulated with the angle $\xi$.

For general values of $\theta_L$ and $\theta_R$, charge 
(\ref{I_charge}) and spin (\ref{I_spin}) Josephson currents 
may be parametrized as $I_J^c = J_1^c \sin (\psi + \chi_1) + J_2^c
\sin (\psi + \chi_2)$ and $I_J^s = J^s \cos (\psi +\chi_1)$, where
$J_1^c (e{\cal F}/G_N) = {\cal C} \delta \cos\xi $, 
$J_2^c (e{\cal F}/G_N) = {\cal D}$,
$J^s (e^2{\cal F}/G_N) = {\cal C} \delta \kappa \sin \xi $, 
$\chi_1 = \tan^{-1}(\frac{ {\cal A}_3 +{\cal A}_4}{{\cal A}_1 +{\cal A}_2})$
and $\chi_2 = \tan^{-1}(\frac{ {\cal A}_3 -{\cal A}_4}{{\cal A}_1 -{\cal A}_2})$
with
${\cal C} = [({\cal A}_1 + {\cal A}_2)^2
+({\cal A}_3 + {\cal A}_4)^2]^{1/2}$, and
${\cal D} = [({\cal A}_1 - {\cal A}_2)^2
+({\cal A}_3 - {\cal A}_4)^2]^{1/2}$.
With the application of external flux, current and phase relationship
\cite{Golubov,Tsuei} for $I_J^c$ may be found out. This will
determine $J_1^c$ and $J_2^c$ and thereby $\delta$ and $\xi$. These
determinations will predict the value of $I_J^s$. The spin current
may be tested if the proposed set-up acts as a source of spin current.

The spin-Josephson current (\ref{I_spin})
is proportional to $\sin \xi$ which means $I_J^s$ is
maximum when the orientation of the axes of non-centrosymmetry between left
and right side of the superconductors are transverse, and it vanishes when the
axes are parallel. This particular property should be useful to control 
spin-Josephson current by orienting the axes about which the inversion symmetries
are lost. Therefore it will be useful to build a {\it Josephson spin switch.}

To summarize, we have shown that the Josephson tunneling process between
two superconductors without inversion symmetry consists of tunneling of
both spin-singlet and spin-triplet Cooper pairs. Because of the direction
dependent tunneling matrix element in real junction and the momentum
dependent phases of the triplet components of the gap function, 
{\it viz.} $\Delta_{\k ,\ua\ua}$ and $\Delta_{\k ,\da\da}$, are opposite,
there is a net spin-Josephson current. This spin current is proportional to
sine of the angle $\xi$ between the axes of non-centrosymmetry in the bulk 
superconductors in each side of the junction. It may be used to make a 
{\it Josephson spin switch} with the controlling parameter $\xi$.

SPM thanks CSIR, Government of India, for his research fellowship.

\end{document}